\documentclass[floats,floatfix,showpacs,amssymb,prd,twocolumn,superscriptaddress,nofootinbib]{revtex4-1}

\usepackage{anyfontsize}
\usepackage{amssymb}
\usepackage{amsmath}
\usepackage{amsfonts}
\usepackage{amsbsy}
\usepackage{float}
\usepackage{txfonts}

\usepackage{subfigure}
\usepackage{gensymb}
\usepackage{soul}
\usepackage{wrapfig}

\def\namedlabel#1#2{\begingroup
    #2%
    \def\@currentlabel{#2}%
    \phantomsection\label{#1}\endgroup
}

\usepackage{mathrsfs}

\usepackage{bm}

\newcommand{\apjl}{Astrophys. J. Lett.}
\newcommand{\mnras}{Mon. Not. R. Astron. Soc.}

\newcommand{\aap}{Astron. \& Astrophys.}

\newcommand{\actaa}{Acta Astronomica}
\newcommand{\jcap}{J. Cosmo. \& Astropart. Phys.}
\newcommand{\apss}{Astrophys. \& Space Sci.}

\newcommand{\mG}{\mathcal{G}}

\usepackage[utf8]{inputenc}

\usepackage{graphicx}
\usepackage{epsfig}
\usepackage{epstopdf}

\usepackage[usenames,dvipsnames]{xcolor}

\usepackage{verbatim,mathtools,needspace,enumitem,etoolbox}%,physics,microtype}

\definecolor{linkcolor}{rgb}{0.0,0.3,0.5}

\usepackage[unicode, colorlinks=true, linkcolor=linkcolor, citecolor=linkcolor, filecolor=linkcolor,urlcolor=linkcolor, pdfusetitle]{hyperref}

\usepackage{epstopdf}

\usepackage{bm}
\usepackage{dcolumn}

\usepackage{longtable}
 \usepackage[normalem]{ulem}

\definecolor{romared}{RGB}{142,0,28}
\hypersetup{colorlinks=true,
            citecolor=romared,
            linkcolor=romared,
            urlcolor=romared}
\makeatletter
\newcommand*{\textoverline}[1]{$\overline{\hbox{#1}}\m@th$}
\makeatother

\begin{document}
%\citestyle{aa}

\title{A 5\% measurement of the gravitational constant in the Large Magellanic Cloud }
% \title{A 5\% constraint on the strength of gravity in the Large Magellanic Cloud}

\author{Harry Desmond}
\email{harry.desmond@physics.ox.ac.uk}
\affiliation{Astrophysics, University of Oxford, Denys Wilkinson Building, Keble Road, Oxford OX1 3RH, UK}

\author{Jeremy Sakstein}
\email{sakstein@hawaii.edu}
\affiliation{Department of Physics \& Astronomy, University of Hawai'i, 2505 Correa Road, Honolulu, Hawai'i, 96822, USA}

\author{Bhuvnesh Jain}
\email{bjain@physics.upenn.edu}
\affiliation{Center for Particle Cosmology,
Department of Physics and Astronomy,
University of Pennsylvania,
%209 S. 33rd St., 
Philadelphia, PA 19104, USA}

\raggedbottom

\begin{abstract}
We perform a novel test of general relativity by measuring the gravitational constant in the Large Magellanic Cloud (LMC). The LMC contains six well-studied Cepheid variable stars in detached eclipsing binaries. Radial velocity {and photometric observations} enable a complete orbital solution, and precise measurements of the Cepheids' periods permit detailed stellar modelling. Both are sensitive to the strength of gravity, the former via Kepler's third law and the latter through the gravitational free-fall time. We jointly fit the observables for stellar parameters and the gravitational constant. Performing a full Markov Chain Monte Carlo analysis of the parameter space including all relevant nuisance parameters, we constrain the gravitational constant in the Large Magellanic Cloud relative to the Solar System to be $G_\text{LMC}/G_\text{SS} = 0.93^{+0.05}_{-0.04}$. We discuss the implications of this 5\% measurement of Newton's constant in another galaxy for dark energy and modified gravity theories. This result excludes one Cepheid, CEP-1812, which is an outlier and needs further study: it is either a highly unusual system to which our model does not apply, or it prefers $G_\text{LMC}<G_\text{SS}$ at $2.6\sigma$. We also obtain new bounds on critical parameters that appear in semi-analytic descriptions of stellar processes. In particular, we measure the mixing length parameter to be $\alpha=0.90^{+0.36}_{-0.26}$ (when assumed to be constant across our sample), and obtain constraints on the parameters describing turbulent dissipation and convective flux. 
\end{abstract}

\date{\today}

\maketitle

\section{Introduction}
\label{sec:intro}

The gravitational constant $G$ is a crucial quantity in physics, underpinning most astrophysical and cosmological studies. Although it is typically assumed to be spatially and temporally constant, this assumption has not been thoroughly tested. Indeed, extra-galactic tests of general relativity constrain post-Newtonian effects such as the Shapiro time-delay \cite{Bolton, Collett:2018gpf} or the existence of new non-Newtonian fifth forces \cite{Burrage:2017qrf,Sakstein:2018fwz,Baker:2019gxo} under the assumption that the baseline value of $G$ in other galaxies is identical to that measured in the Solar System. A possible cosmological time-variation of $G$ has been constrained using Solar System \cite{Williams:2004qba,Hofmann_2018,mars}, stellar \cite{GarciaBerro:2011wc,Mould:2014iga,Bellinger:2019lnl}, pulsar timing \cite{Zhu:2018etc}, and cosmological \cite{WiggleZ, Ooba:2017gyn,Alvey:2019ctk} measurements. The spatial variation of $G$ is poorly constrained despite being a widespread prediction of dark energy models \cite{Koyama:2015oma,Sakstein:2015zoa,Crisostomi:2019yfo,Sakstein:2018fwz,Baker:2019gxo}, as well as potential resolutions of the Hubble tension \cite{D19, rho_DM, D20}. Large scale structure tests on {$\sim$1$-200$} Mpc scales and $0.3\lesssim z \lesssim 1$ (e.g.~\cite{Simpson, DES_MGconstraints, Joudaki_MG, Johnson, Quintero}) have uncertainties at the $\ge$20\% level, and leave open the possibility of variation on galaxy scales and below. %, even beyond the relatively large uncertainties ($\sim$20$-50\%$). BJ: I removed this phrase since it seemed to risk confusing the reader.
See Ref.~\cite{Will} for a review of Solar System tests of general relativity, \cite{Burrage:2017qrf} for a review of laboratory tests, \cite{Sakstein:2018fwz,Baker:2019gxo} for reviews of astrophysical tests, and \cite{Koyama:2015vza,Ferreira:2019xrr} for reviews of cosmological tests.

We present a novel test of the spatial variation of $G$ and demonstrate its potential by applying it to observations of stars in the Large Magellanic Cloud (LMC). Consider a detached eclipsing binary (DEB) composed of a variable star and its companion. Radial velocity {and photometric} measurements can be used to construct an orbital solution, which allows for the determination of the stellar masses of both components and the cycle-averaged radius of the variable star. Simultaneously, modelling the light curves allows parameters of the variable star to be determined independently \cite{Marconi:2013tta}. Both of these measurements depend on $G$.

The orbital solution is derived using Kepler's third law
\begin{equation}
    \frac{4\pi^2 a^3}{\Pi^2}=G(M+m),
\end{equation}
where $a$ is the semi-major axis of the binary, $\Pi$ is its period, $M$ is the mass of the variable star, and $m$ the mass of its companion. $GM$ and $Gm$ may be separately inferred in DEBs from simultaneous measurement of the orbital period and the orbital radii of the two stars. The inference of $m/M$ is independent of $G$, allowing us to focus attention on $GM$. The variable star's pulsation period is approximately proportional to $(GM)^{-1/2}$ (the free-fall time), although a detailed numerical solution yields slightly different powers of $G$ and $M$ \cite{Cox}. We will show that consistency of stellar parameters from orbital and pulsational modelling has the potential to place strong bounds on $G$. Note that in general only dimensionless quantities can be measured \cite{Dicke, Scott, Uzan}, so it is important to quote variations in $G$ in dimensionless terms. We focus here on the ratio of $G$ in the LMC to that inferred in the Solar System, $G_\text{LMC}/G_\text{SS}$.

The LMC contains a population of Cepheid variable stars in DEBs \cite{Araucaria}. This population is ideal to demonstrate the power of our proposed test since Cepheids are well-understood objects that can be modelled numerically to high precision. Indeed, these objects have been useful in testing fundamental physics previously \cite{Jain, D19}. Ancillary LMC observations such as the location of the tip of the red giant branch (TRGB) allow us to break degeneracies by providing a distance to the LMC that can be used to convert apparent magnitudes to luminosities. The rest of this paper explains our procedure for modelling these objects and applies it to the LMC Cepheid DEBs. We find sufficient constraining power in the data to measure $G$ to $5\%$.

The paper is structured as follows. In Sec.~\ref{sec:method} we describe the data that we use, document our model for stellar structure, and present the likelihood framework within which we derive constraints. Sec.~\ref{sec:results} presents our results, while Sec.~\ref{sec:discussion} discusses the significance of the results, considers systematic uncertainties and proposes future work. Sec.~\ref{sec:conc} concludes.

\section{Methodology}
\label{sec:method}

\subsection{Observational data}
\label{sec:data}

Our data comes from the Araucaria Project \cite{Araucaria}, a study of seven Cepheids in the LMC that form part of DEB systems. These were selected because they form a well-studied sample, and because their location in the LMC will allow us to constrain $G$ beyond the Milky Way's disk. The measurements derive primarily from the Optical Gravitational Lensing Experiment (OGLE) \cite{OGLE}, although all the objects were separately followed up spectroscopically. We refer the reader to Ref.~\cite{Araucaria} and references therein for a full description of the project and the methods employed.

We neglect LMC-T2CEP-098 since it has more in common with Type-II than Type-I Cepheids \cite{Araucaria}, and our stellar structure modelling (see below) is unable to treat these objects reliably at this time \cite{MESA_RSP}. For each of the remaining six Type-I Cepheids, we use the measurements of mass $M$, pulsation period $P$, cycle-averaged radius $R$, and $V$- and $I$-band magnitudes (Table~\ref{tab:cephprops}; see also Ref.~\cite{Araucaria} tables 1, 3, 5, 7, 9 \& 11).  $V$, $I$, $P$ and $R$ are measured directly independently of $G$, the first three from the photometric flux and its time-variation and the latter from the eclipse curve of the binary. However, $M$ is inferred from modelling the orbits of the stars assuming Newtonian mechanics (i.e. the same value for the gravitational constant as measured locally, $G_\text{LMC}=G_{\rm SS}$). In particular it is the product $GM$ that appears in Kepler's laws. Thus while we can compare our model to $V$, $I$, $P$ and $R$ directly we must modify the $M$ constraint according to the value of $G$ we are testing, as described in Sec.~\ref{sec:likelihood}. The luminosities and temperatures inferred by Ref.~\cite{Araucaria} depend in a non-trivial way on $G$, so we do not use these constraints in our analysis: these are input parameters to our models that we will constrain independently of Ref.~\cite{Araucaria}.

\begin{table*}
  \begin{center}
  \small\addtolength{\tabcolsep}{-5pt}
    \begin{tabular}{|l|c|c|c|c|c|c|c|c|c|c|}
      \hline
      Cepheid label & $G/G_\text{SS} \; M/{\rm M}_\odot$ & $\Delta M/{\rm M}_\odot$ & $P$/day & $\Delta P$/day & $R/R_\odot$ & $\Delta R/R_\odot$ & $V$ & $\Delta V$ & $I$ & $\Delta I$\\ 
      \hline
    \rule{0pt}{3.5ex}
      CEP-1718a & 4.27 & 0.04 & 1.9636625 & 0.0000063 & 27.80 & 1.20 & 15.72 & 0.03 & 15.21 & 0.036\\
    \rule{0pt}{3.5ex}
      CEP-1718b & 4.22 & 0.04 & 2.4808680 & 0.0000116 & 33.10 & 1.30 & 15.74 & 0.03 & 15.22 & 0.036\\
    \rule{0pt}{3.5ex}
      CEP-1812 & 3.76 & 0.03 & 1.3129039 & 0.0000009 & 17.85 & 0.13 & 16.87 & 0.07 & 16.30 & 0.076\\
    \rule{0pt}{3.5ex}
      CEP-0227 & 4.15 & 0.03 & 3.7970782 & 0.0000114 & 34.87 & 0.12 & 15.55 & 0.04 & 14.96 & 0.064\\
    \rule{0pt}{3.5ex}
      CEP-2532 & 3.98 & 0.10 & 2.0353516 & 0.0000103 & 29.20 & 1.40 & 15.67 & 0.05 & 15.17 & 0.054\\
    \rule{0pt}{3.5ex}
      CEP-4506 & 3.61 & 0.03 & 2.9878239 & 0.0000075 & 28.50 & 0.20 & 15.78 & 0.03 & 15.21 & 0.050\\
      \hline
    \end{tabular}
  \caption{Properties of the LMC Cepheids used in our inference. The data are taken from Ref.~\cite{Araucaria} and the OGLE database.\footnote{\url{http://ogledb.astrouw.edu.pl/~ogle/OCVS/}}}
  \label{tab:cephprops}
  \end{center}
\end{table*}

\subsection{Simulating stellar structure}
\label{sec:mesa}

To model the internal properties of the Cepheids we use the stellar structure code \texttt{MESA} (Modules for Experiments in Stellar Astrophysics, version 12266; \cite{MESA_1, MESA_2}). In particular, we utilise the Radial Stellar Pulsation (RSP) module \cite{MESA_RSP}, which incorporates the pulsation code of Ref.~\cite{Smolec:2008qh}. We have modified the code to allow for a variable value of Newton's constant. Given a set of input parameters as described below, RSP creates an initial stellar model which is evolved through a series of radial pulsations until convergence. The module includes eight parameters that control the efficiencies of various convective and turbulent processes. These strongly affect the shape and amplitude of the light curves, but only three of them---$\alpha_\text{d}$ (turbulent dissipation), $\alpha_\text{c}$ (convective flux), and $\alpha$ (mixing length)---{significantly} affect the period. For example, varying $\alpha_m$ (describing eddy-viscous dissipation) across its expected range in Ref.~\cite{MESA_RSP} produces a $5\times$ smaller change to the period than do $\alpha_\text{d}$, $\alpha_\text{c}$ or $\alpha$. We treat these three as nuisance parameters and marginalise over them. We do not use the shape or amplitude of the light curves in our inference due to their high sensitivity to all eight nuisance parameters, so we fix the remaining five parameters to their fiducial values in Ref.~\cite{MESA_RSP}. (See Ref.~\cite{Smolec:2008qh} for a full description of the nuisance parameters in \texttt{MESA}.) We fix the atmosphere model, which sets the boundary conditions for the pulsation equations, to the \texttt{MESA} default. The choice of boundary conditions may therefore be a potential source of systematic error  \cite{2018ApJ...860..131C}.

The free parameters we use for each star are mass $M$, bolometric luminosity $L$, effective temperature $T$, hydrogen mass fraction $X$, and metallicity $Z$, in addition to the strength of gravity $G$, and the nuisance parameters $\alpha_\text{d}$, $\alpha_\text{c}$ and $\alpha$. {In order for the joint inference of this many parameters to be computationally tractable, we cannot run a full \texttt{MESA} simulation (which takes $\sim$4$-12$ hours) in each likelihood evaluation.} We therefore create fitting formulae for the final converged period and radius values $\{P_1, R_1\}$ as a function of the linear values $\{P_0, R_0\}$ that are calculated following the creation of the initial model. This is done separately for each star by performing full \texttt{MESA} simulations for a set of points in parameter space surrounding the best-fit values of Ref.~\cite{Araucaria}, with $G$ in the range $(0.8-1.2) \, G_\text{SS}$. Removing outlier cases in which the model fails to converge, we find
\begin{align}\label{eq:correction}
P_1 &= \beta_P \: P_0 + \gamma_P \: P_0^2\\ \nonumber
R_1 &= \beta_R \: R_0 + \gamma_R \: R_0^2,
\end{align}
with fractional Gaussian scatter in $P_1$ and $R_1$ given by $\sigma_P$ and $\sigma_R$ respectively. The fit parameters for the 6 stars are given in Table~\ref{tab:fitparams}. Henceforth we will use `$P$' and `$R$' for $P_1$ and $R_1$ in Eq.~\eqref{eq:correction}, with theoretical uncertainties given by the scatters in the fit relations. These typically exceed the measurement uncertainties for $P$ but not for $R$. Note that the periods are specific to the mode that the Cepheids pulsate in, either the fundamental (CEP-1812, CEP-0227 and CEP-4506) or the first overtone (CEP-1718a,b and CEP-2532).

\begin{table*}
  \begin{center}
  \small\addtolength{\tabcolsep}{-5pt}
    \begin{tabular}{|l|c|c|c|c|c|c|}
      \hline
      Cepheid label & $\beta_P$ & $\gamma_P$ & $\sigma_P$ & $\beta_R$ & $\gamma_R$ & $\sigma_R$\\ 
      \hline
    \rule{0pt}{3.5ex}
      CEP-1718a & 1.000024 & $4.91 \times 10^{-5}$ & $1.15 \times 10^{-5}$ & 0.9999417 & $3.15 \times 10^{-6}$ & $8.08 \times 10^{-5}$\\
    \rule{0pt}{3.5ex}
      CEP-1718b & 1.000049 & $3.80 \times 10^{-5}$ & $1.04 \times 10^{-5}$ & 0.9999538 & $1.56 \times 10^{-6}$ & $1.36 \times 10^{-5}$\\
    \rule{0pt}{3.5ex}
      CEP-1812 & 1.000085 & $-1.53 \times 10^{-6}$ & $7.61 \times 10^{-6}$ & 1.0000000 & $1.76 \times 10^{-9}$ & $1.18 \times 10^{-8}$\\
    \rule{0pt}{3.5ex}
      CEP-0227 & 1.000083 & $1.70 \times 10^{-6}$ & $4.58 \times 10^{-6}$ & 0.9999997 & $1.08 \times 10^{-8}$ & $4.89 \times 10^{-8}$\\
    \rule{0pt}{3.5ex}
      CEP-2532 & 1.000036 & $4.85 \times 10^{-5}$ & $1.30 \times 10^{-5}$  & 0.9999892 & $5.50 \times 10^{-7}$ & $8.99 \times 10^{-6}$\\
    \rule{0pt}{3.5ex}
      CEP-4506 & 1.000079 & $3.65 \times 10^{-6}$ & $5.18 \times 10^{-6}$ & 0.9999999 & $6.91 \times 10^{-9}$ & $7.37 \times 10^{-8}$\\
      \hline
    \end{tabular}
  \caption{Parameters for the fitting functions relating linear period and radius to their converged values (Eq.~\eqref{eq:correction}).}
  \label{tab:fitparams}
  \end{center}
\end{table*}

We have made several other modifications to \texttt{MESA} both to reduce the runtime and to facilitate the running of multiple \texttt{MESA} instances simultaneously within an MPI- and OpenMP-parallelised Markov Chain Monte Carlo (MCMC) framework. We make our modifications to \texttt{MESA} and likelihood code publicly available.\footnote{\url{https://zenodo.org/record/4309065 } \cite{harry_desmond_2020_4309065}}

\subsection{Likelihood model}
\label{sec:likelihood}

Our likelihood has four contributions for each Cepheid, describing the comparison of the model to the measured masses, luminosities, periods, and radii respectively. The first describes the constraint on the dynamical mass deriving from the orbital solution for the binary system. Although quoted as a bound on $M$ in Ref.~\cite{Araucaria}, this assumes $G_\text{LMC}=G_{\rm SS}$. Since it is the product $GM$ that appears in Kepler's laws, allowing for  variable $G$ the quoted constraint is really on $G_{\rm LMC}M/G_{\rm SS}$. Thus for e.g. $G_\text{LMC} < G_\text{SS}$, $M$ must be larger to achieve the same orbital fit. The log-likelihood therefore has the component
\begin{equation}\label{eq:Lm}
\ln(\mathcal{L}_{M,i}) = -\frac{1}{2} \left(\ln(2 \pi \Delta M_{d,i}^2) + (\mathcal{G}_i M_i - M_{d,i})^2/\Delta M_{d,i}^2\right),
\end{equation}
where $i$ labels the Cepheid, $d$ denotes observed value (see Table~\ref{tab:cephprops}), and $\mathcal{G}_i \equiv G_i/G_\text{SS}$.

The next contribution derives from the Cepheids' magnitudes. Since the $V$- and $I$-band fluxes of the Cepheids are measured, knowledge of their distance translates to knowledge of their luminosities, which are input quantities to \texttt{MESA}. We calculate the uncertainty-weighted best-fit value of the distance to the LMC measured by the TRGB method from the \textit{NASA Extragalactic Distance Database} (NED-D)\footnote{\url{https://ned.ipac.caltech.edu/Library/Distances/}} \cite{NED-D} as $50.5 \pm 0.3$ kpc. Since we do not know where the Cepheids are within the LMC along the line of sight, we include an additional uncertainty of 4.4 kpc corresponding to the approximate radius of the LMC (5\degree). This gives $\Delta D_\text{LMC}=4.41$ kpc. The TRGB measurement does however depend on $G$, so the distance must be corrected if $G_\text{LMC} \ne G_{\rm SS}$. This correction was calculated in Refs.~\cite{rho_DM, D19}:
\begin{equation}\label{eq:D_LMC}
D_\text{LMC}(\mG) = 1.021 \: (1 - 0.04663 \: \mG^{8.389})^{1/2} \: D_\text{LMC}(\mG=1).
\end{equation}
Note that we assume an independent $G$ for each star when evaluating Eq.~\ref{eq:D_LMC}. Although it may be preferable to use a combined $G_\text{LMC}$, coupling the stars, the modification to $D_\text{LMC}$ from this equation is subdominant to $\Delta D_\text{LMC}$ for values of $G$ across our posteriors. We also caution that this relation was derived at fixed mass and metallicity, so may be subject to small uncertainties from their variation. Another source of theoretical error not accounted for is uncertainty in the triple-$\alpha$ rate. This however is expected to be subdominant to the other uncertainties \cite{2017A&A...606A..33S}.

We use this distance to convert the $V$- and $I$-band magnitudes quoted in Ref.~\cite{Araucaria} to absolute magnitudes 
and their uncertainties. We then convert these to bolometric luminosities using bolometric corrections computed by \texttt{MESA}, and take the uncertainty-weighted average of the $V$- and $I$-band results and its uncertainty as our estimates of $M_\text{bol}$ and $\Delta M_\text{bol}$. 
% We then calculate the luminosity as
Finally, we compare the associated $L_\text{bol}$ and $\Delta L_\text{bol}$ to the assumed bolometric luminosity $L$ (as input to \texttt{MESA}) in the likelihood term
\begin{equation}\label{eq:Ll}
\ln(\mathcal{L}_{L,i}) = -\frac{1}{2} \left(\ln(2 \pi \Delta L_\text{bol,i}^2) + (L_i - L_\text{bol,i})^2/\Delta L_\text{bol,i}^2\right).
\end{equation}

The third and fourth components of the likelihood describe the Cepheids' pulsation periods and radii. Both of these quantities are measured in Ref.~\cite{Araucaria}, and as the measurement methods are insensitive to $G$ we can import them directly into our analysis. Using the predicted periods $P$ and radii $R$ described in Sec.~\ref{sec:mesa}, we have
\begin{align}
\label{eq:L_P} \ln(\mathcal{L}_{P,i}) &= -\frac{1}{2} \left(\ln(2 \pi \Delta P_\text{tot,i}^2) + (P_i - P_\text{d,i})^2/\Delta P_\text{tot,i}^2\right),\\
\ln(\mathcal{L}_{R,i}) &= -\frac{1}{2} \left(\ln(2 \pi \Delta R_\text{tot,i}^2) + (R_i - R_\text{d,i})^2/\Delta R_\text{tot,i}^2\right),
\end{align}
where $\Delta P_\text{tot,i}$ and $\Delta R_\text{tot,i}$ are the quadrature sums of the theoretical and observational uncertainties.

We assume independence of the $M$, $L$, $P$, and $R$ measurements, so that the overall log-likelihood for star $i$ is given by:
\begin{align}
\ln(\mathcal{L}_i(&M_\text{d}, L_\text{bol}, P_\text{d}, R_\text{d}|G, M, T, L, X, Z, \alpha_\text{d}, \alpha_\text{c}, \alpha)) = \\ \nonumber
&\ln(\mathcal{L}_{M,i}) + \ln(\mathcal{L}_{L,i}) + \ln(\mathcal{L}_{P,i}) + \ln(\mathcal{L}_{R,i}). 
\end{align}
We use Bayes' theorem to derive the posteriors on the model parameters, assuming a uniform prior in the range $1 < M/\text{M}_\odot < 15$, $2000 < T/\text{K} < 9500$, $200 < L/\text{L}_\odot < 2100$, $0.01 < X < 1$, $0.0001 < Z < 0.02$, $0.1 < G_\text{LMC}/G_\text{SS} < 1.44$, $0.01 < \alpha_\text{d} < 3$, $0.01 < \alpha_\text{c} < 3$ and $0.01 < \alpha < 3.5$. These widely bracket the expected values. We perform MCMC with the \textit{PyMultiNest} algorithm \cite{MN1, MN2, pyMN}, inferring the 9 input parameters of each star separately for computational expediency. Each likelihood evaluation takes $\sim$15s for a given star (dominated by the \texttt{MESA} simulation), and convergence of the MCMC {(with \texttt{n\_live\_points}=800, \texttt{importance\_nested\_sampling}=True, \texttt{multimodal}=True, \texttt{evidence\_tolerance}=0.8, \texttt{sampling\_efficiency}=0.4 and \texttt{const\_efficiency\_mode}=False)} takes $\sim$3 weeks on $3\times28$ cores. We also combine the constraints on $G_\text{LMC}/G_\text{SS}$, $\alpha_\text{d}$, $\alpha_\text{c}$ and $\alpha$ from the separate stars, which could plausibly be in common between them. We do this by deriving 1D marginalised posteriors for each of these parameters for each star from the density of samples in the equally-weighted chains. We then use these as likelihood functions in a second MCMC step in which we combine the constraints from all of the stars by multiplying the likelihoods.

\section{Results}
\label{sec:results}

\begin{figure*}
  \centering
  \includegraphics[width=0.44\textwidth]{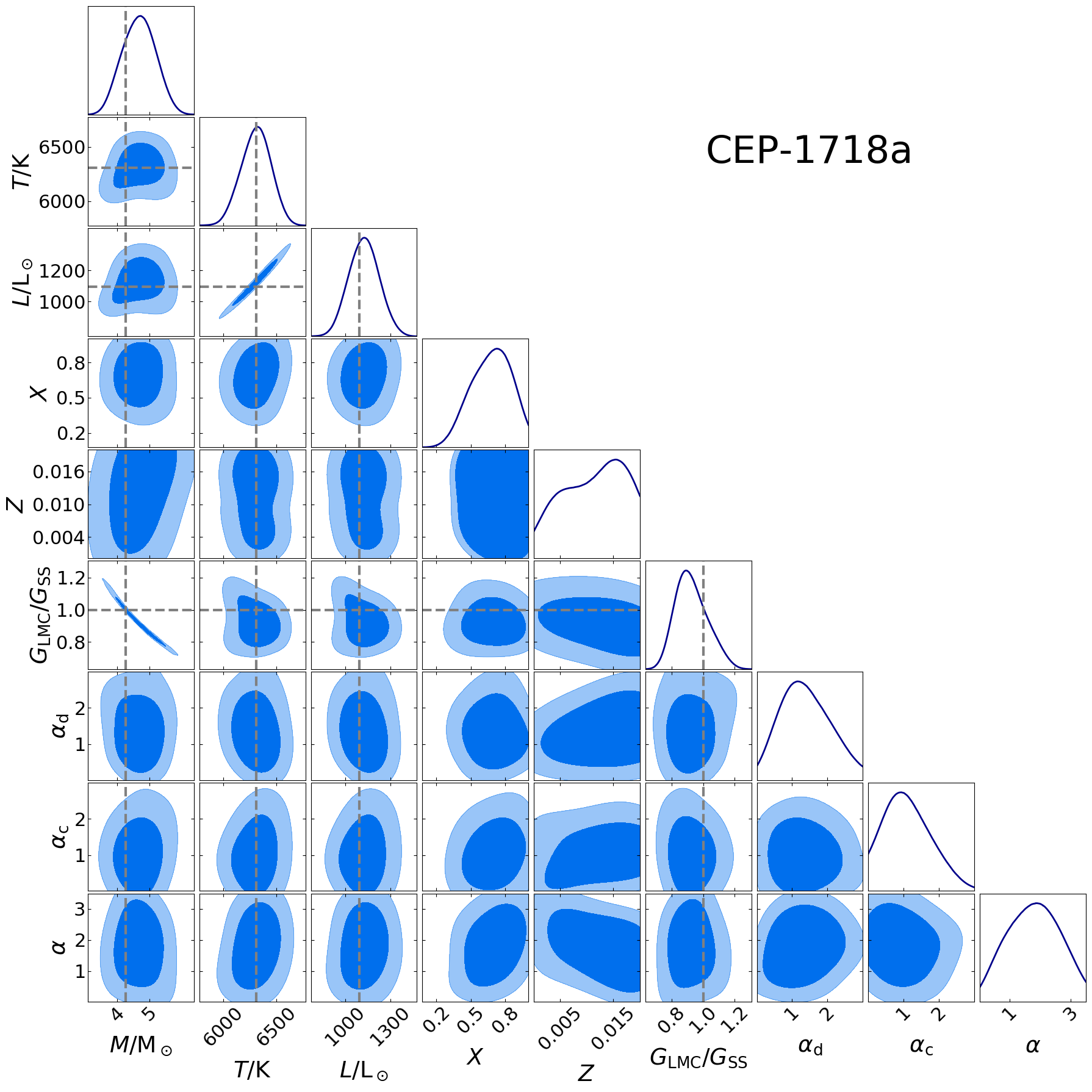}
  \includegraphics[width=0.44\textwidth]{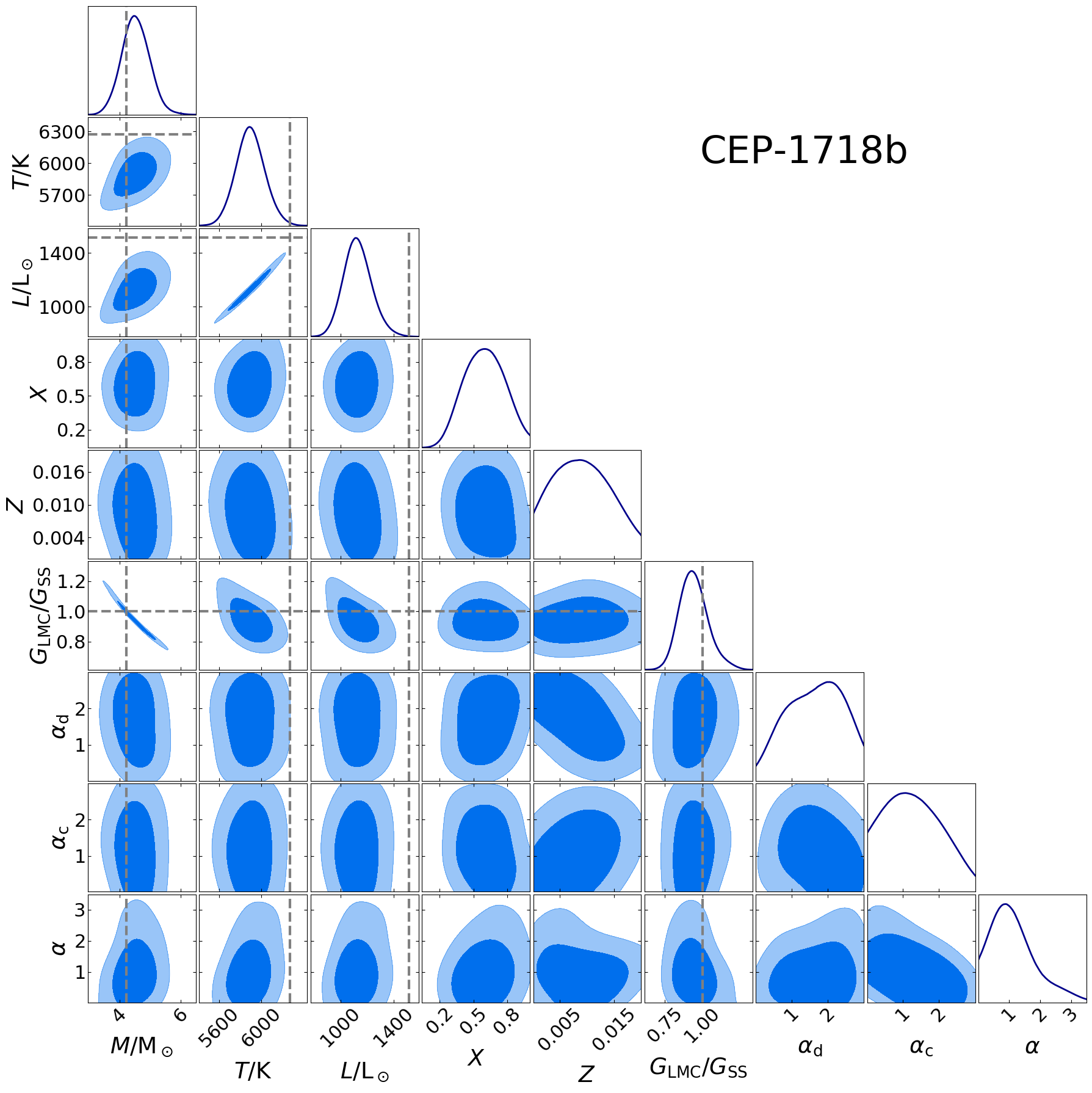}
  \includegraphics[width=0.44\textwidth]{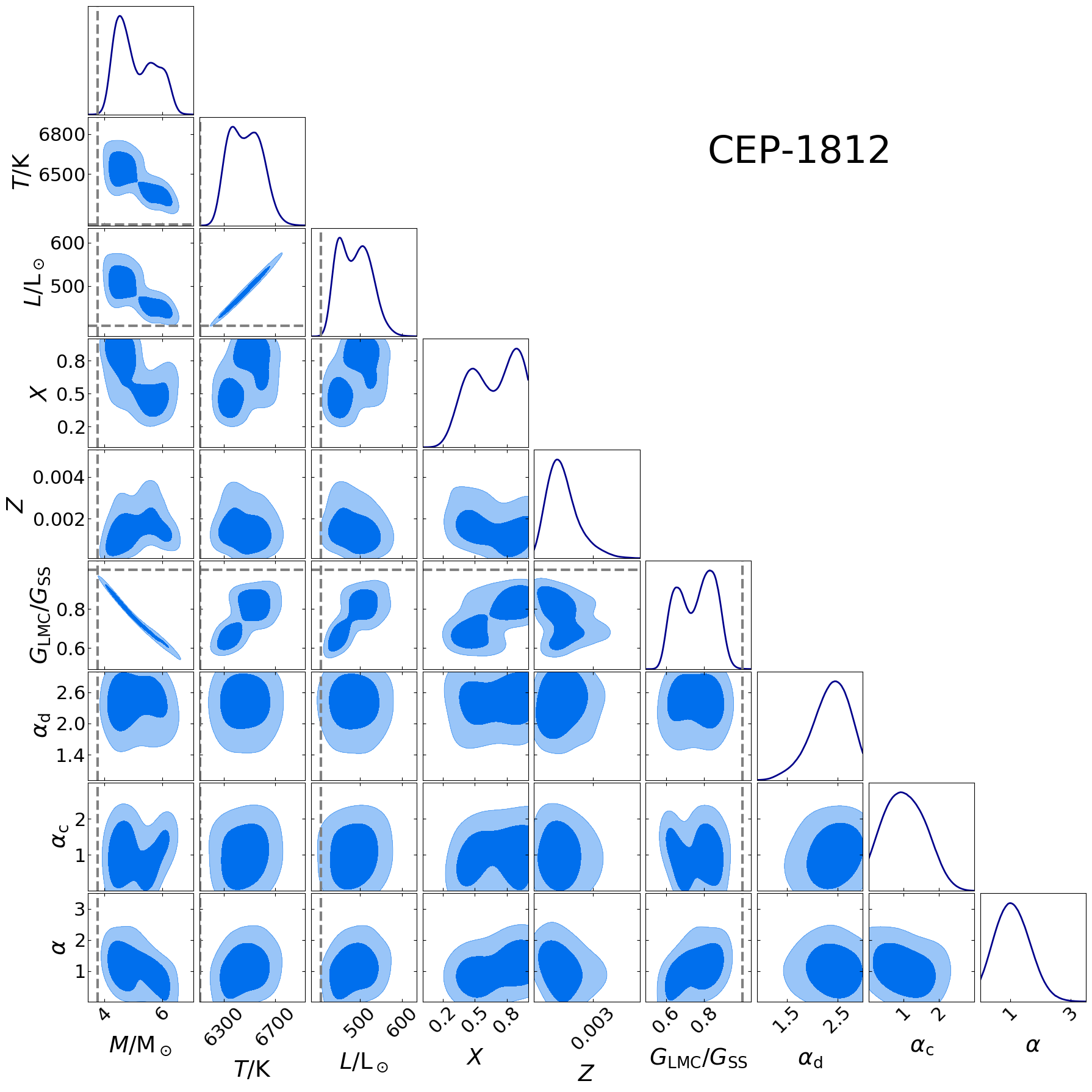}
  \includegraphics[width=0.44\textwidth]{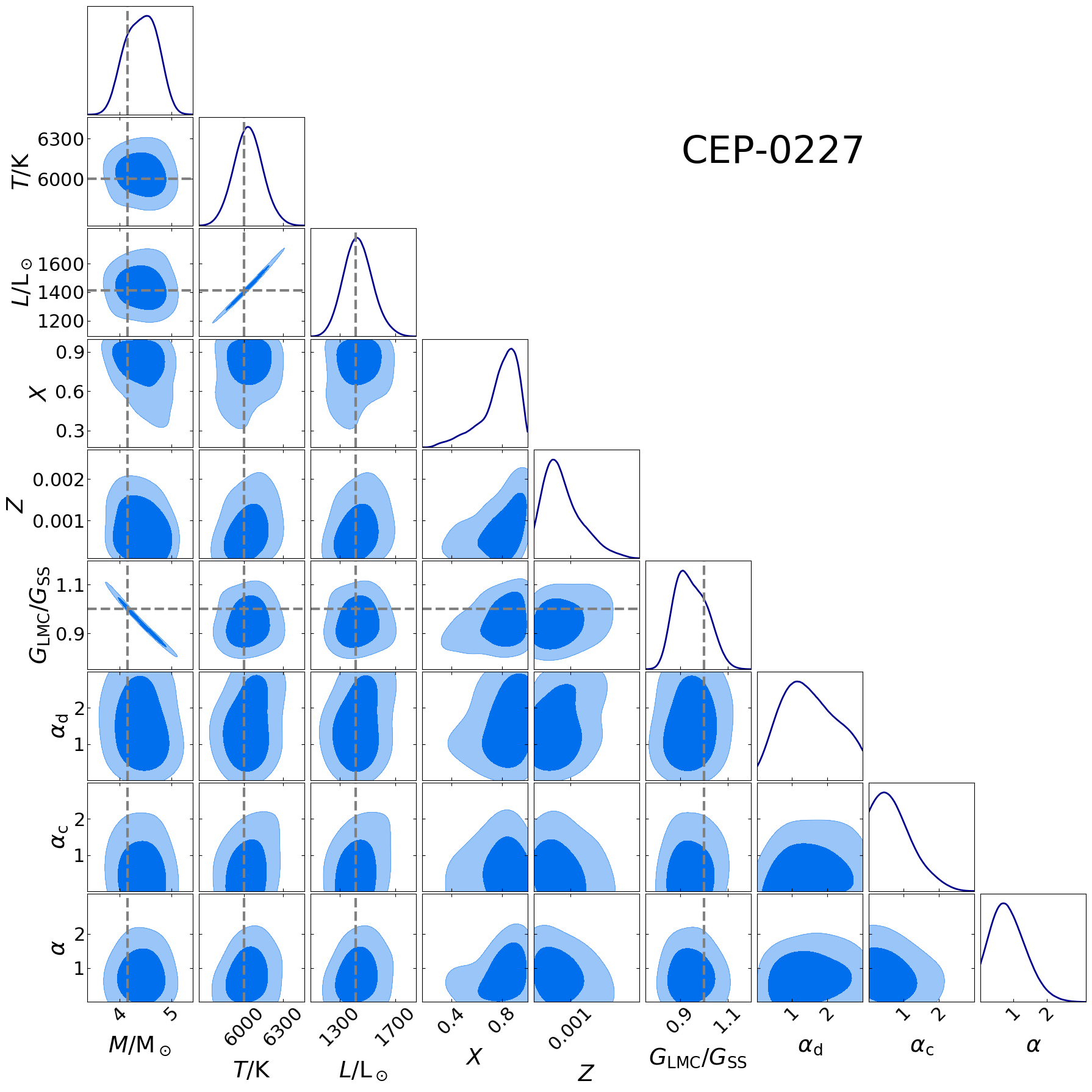}
  \includegraphics[width=0.44\textwidth]{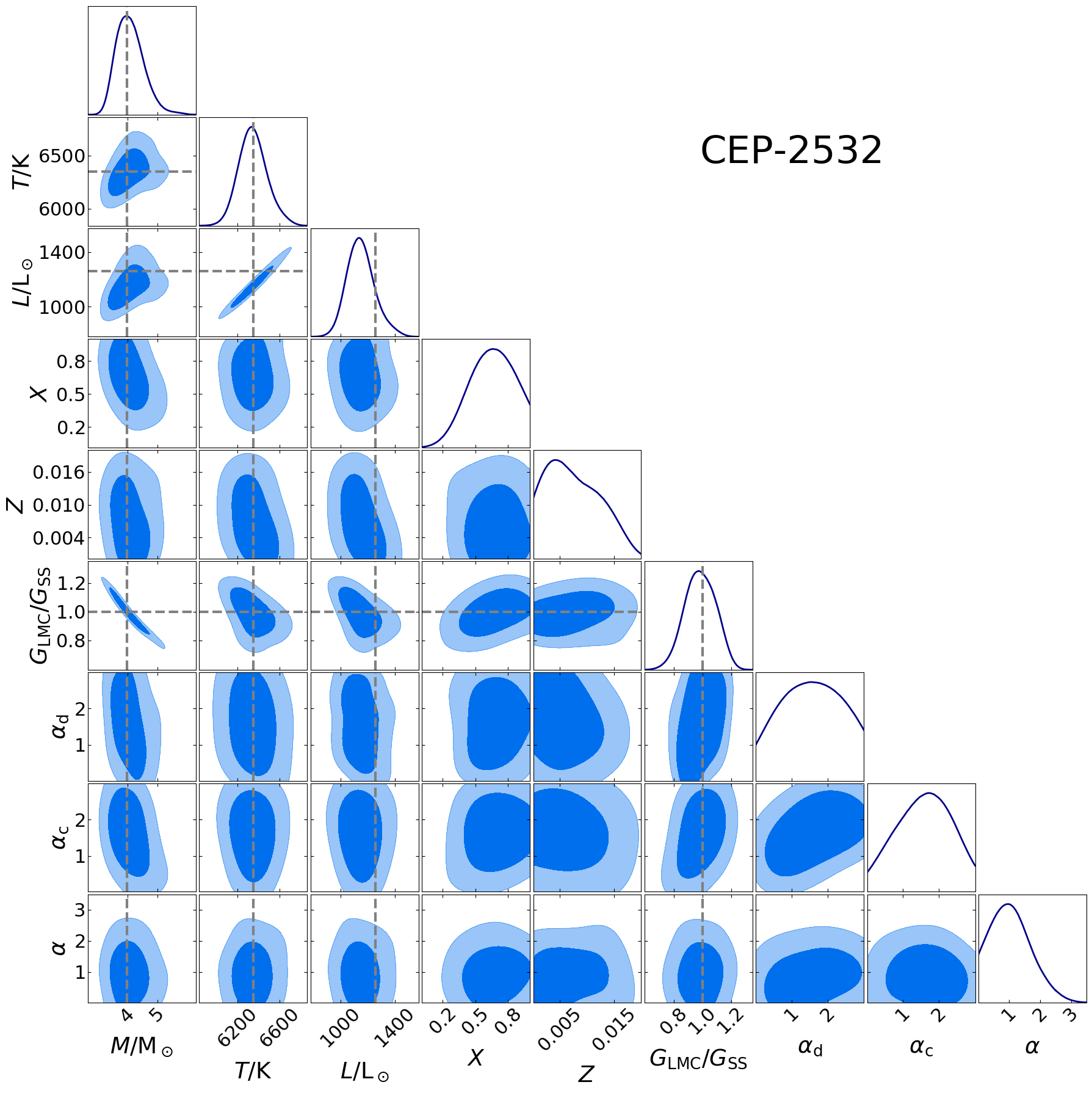}
  \includegraphics[width=0.44\textwidth]{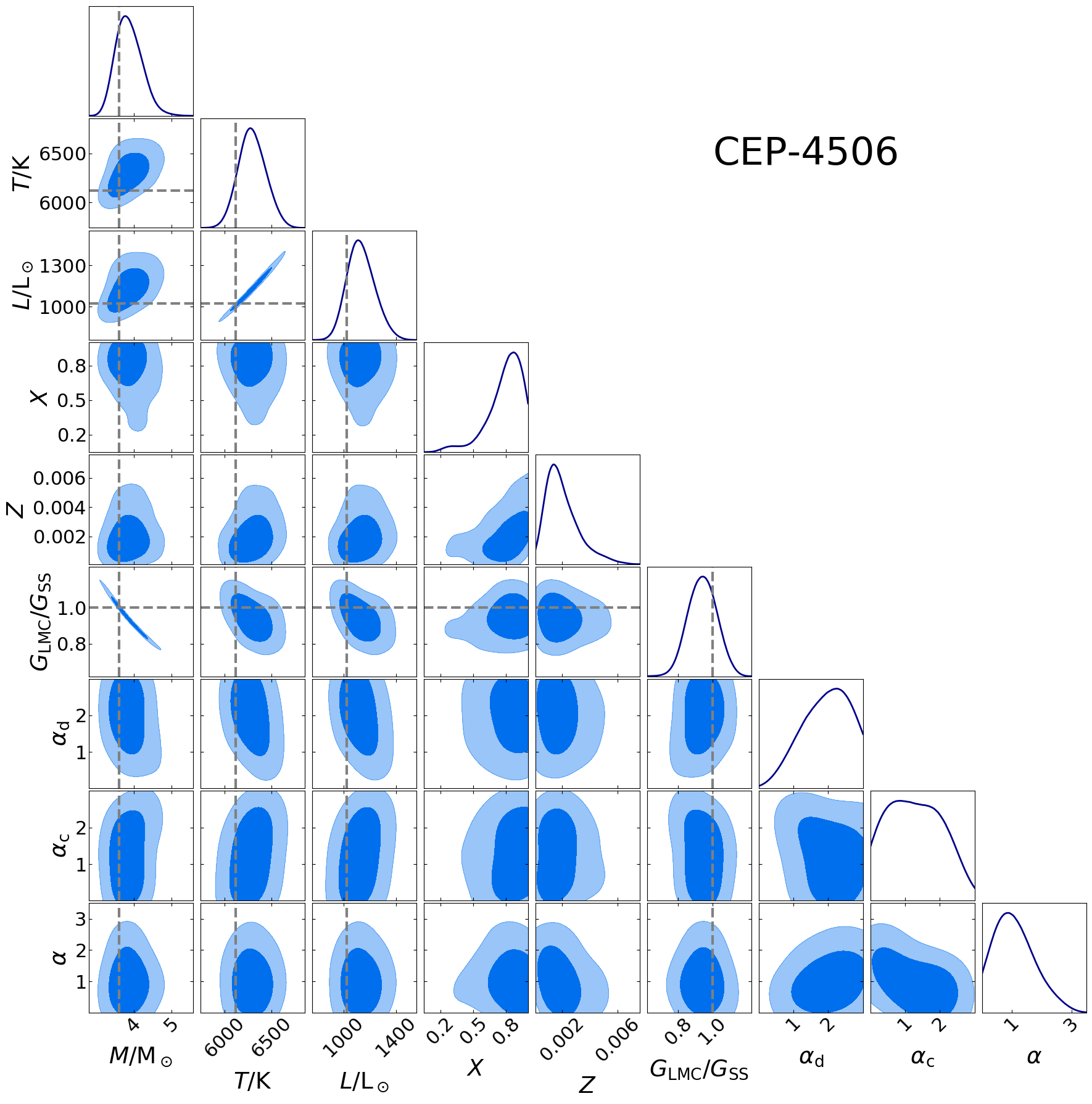}
  \caption{Corner plots of parameter constraints on all 6 Cepheids.}
  \label{fig:posteriors}
\end{figure*}

The full corner plots of the constraints on all 9 parameters for each star are given in Fig.~\ref{fig:posteriors}, while the median values and $1\sigma$ constraints are shown in Table~\ref{tab:results}. The constraints on $G_\text{LMC}/G_\text{SS}$ from the individual stars are $\sim$10\%, while those on $\alpha_\text{d}$, $\alpha_\text{c}$ and $\alpha$ are $\sim$50\%. 
We find that CEP-1812 gives values of $G_\text{LMC}/G_\text{SS}$ and $\alpha_\text{d}$ inconsistent with the other stars at the $\sim$1.5$\sigma$ level. As discussed in Sec.~\ref{sec:errors} this is an unusual star in several ways, and likely violates assumptions used in our analysis. We therefore exclude it from our fiducial joint constraints on $G_\text{LMC}/G_\text{SS}$, $\alpha_\text{d}$, $\alpha_\text{c}$ and $\alpha$ shown in Fig.~\ref{fig:joint}, although we show the results with CEP-1812 included in Appendix~\ref{sec:app}. 

For the remaining 5 stars we obtain: 
\begin{equation}
    \frac{G_{\rm LMC}}{G_{\rm SS}}=0.927^{+0.051}_{-0.037},
\end{equation}
a 4.6\% constraint on $G_\text{LMC}/G_\text{SS}$ consistent with 1 at $1.5\sigma$. To our knowledge this is the first measurement of $G$ in another galaxy, complementing the direct constraints on the parametrised post-Newtonian parameter $\gamma$ that have already been achieved \cite{Bolton, Collett:2018gpf}. We obtain $\sim$35\% constraints on $\alpha_\text{d}$, $\alpha_\text{c}$ and $\alpha$ when these are assumed to be the same between the stars (see first row of Table~\ref{tab:results_joint}). Note that although $\alpha_\text{d}$, $\alpha_\text{c}$ and $\alpha$ are nuisance parameters in the inference of $G_\text{LMC}/G_\text{SS}$, they are of great importance for stellar astrophysics. In particular, the mixing length parameter $\alpha$ determines the location of the Hayashi track and the details of post-main-sequence stellar evolution.

\begin{figure*}
  \centering
  \includegraphics[width=0.49\textwidth]{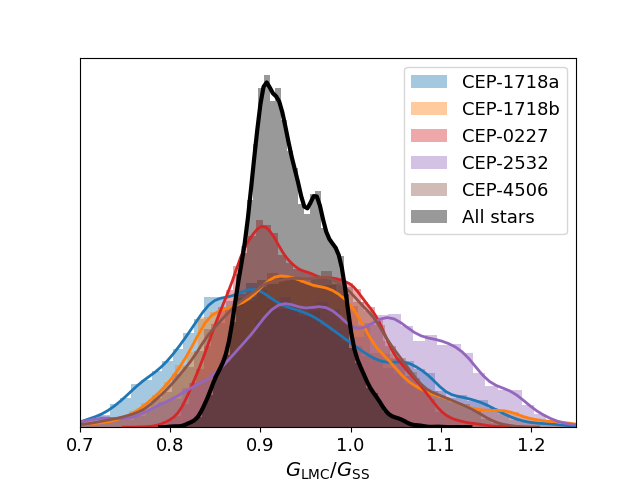}
  \includegraphics[width=0.49\textwidth]{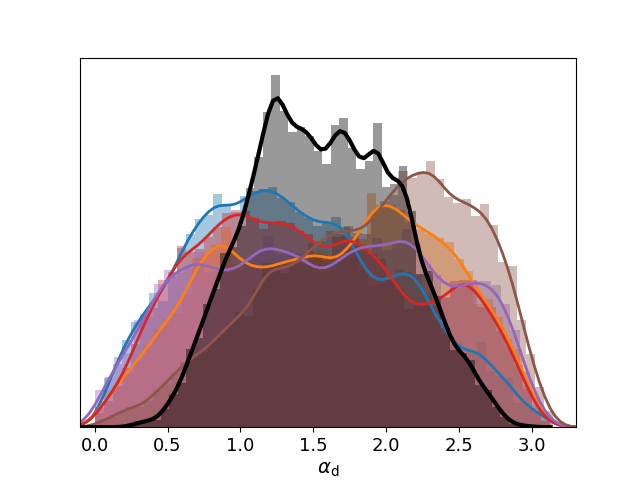}
  \includegraphics[width=0.49\textwidth]{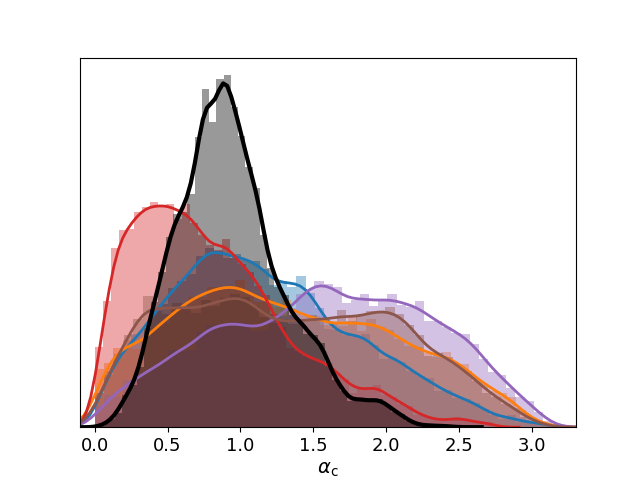}
  \includegraphics[width=0.49\textwidth]{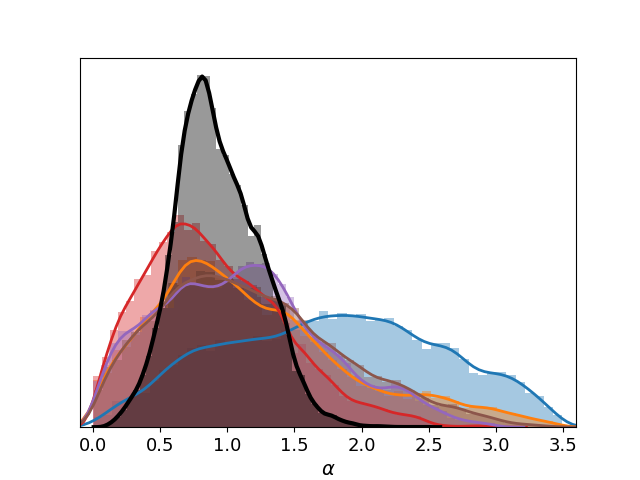}
  \caption{Combined posterior probability distributions for $G_\text{LMC}/G_\text{SS}$, $\alpha_\text{d}$, $\alpha_\text{c}$ and $\alpha$ with CEP-1812 removed.}
  \label{fig:joint}
\end{figure*}

The only significant parameter degeneracies shown in Fig.~\ref{fig:posteriors} are between $M$ and $G$, and $T$ and $L$. The former is due to the $GM$ constraint encapsulated in Eq.~\ref{eq:Lm}, which makes $G$ roughly inversely proportional to $M$. This degeneracy is primarily broken by the period term in the likelihood (Eq.~\ref{eq:L_P}) which we find to be sensitive to relatively small variations in $G$ at fixed $GM$. The $T-L$ degeneracy derives from the Stefan-Boltzmann law $L = 4 \pi \sigma R^2 T^4$ with $R$ very strongly constrained in the input. This degeneracy is tighter for the fundamental mode pulsators for which $R$ has been measured to greater precision.
% The remaining weak degeneracies derive from the complex interplay of the various parameters on a star-by-star basis.

\begin{table*}
  \begin{center}
  \small\addtolength{\tabcolsep}{-5pt}
    \begin{tabular}{|l|c|c|c|c|c|c|c|c|c|}
      \hline
      Cepheid label & $M/{\rm M}_\odot$ & $T/\text{K}$ & $L/{\rm L}_\odot$ & $X$ & $Z$ & $G/G_\text{SS}$ & $\alpha_\text{d}$ & $\alpha_\text{c}$ & $\alpha$\\ 
      \hline
    \rule{0pt}{3.5ex}
      CEP-1718a & $4.69^{+0.50}_{-0.57}$ & $6310^{+130}_{-150}$ & $1130^{+100}_{-100}$ & $0.68^{+0.16}_{-0.20}$ & $0.012^{+0.005}_{-0.007}$ & $0.91^{+0.13}_{-0.09}$ & $1.32^{+0.80}_{-0.63}$ & $1.08^{+0.75}_{-0.57}$ & $1.80^{+0.86}_{-0.94}$\\
    \rule{0pt}{3.5ex}
      CEP-1718b & $4.52^{+0.47}_{-0.42}$ & $5890^{+140}_{-130}$ & $1120^{+100}_{-100}$ & $0.59^{+0.19}_{-0.18}$ & $0.009^{+0.006}_{-0.005}$ & $0.93^{+0.10}_{-0.09}$ & $1.69^{+0.74}_{-0.88}$ & $1.24^{+0.88}_{-0.74}$ & $1.04^{+0.80}_{-0.57}$\\
    \rule{0pt}{3.5ex}
      CEP-1812 & $4.90^{+0.98}_{-0.53}$ & $6460^{+140}_{-130}$ & $488^{+40}_{-41}$ & $0.68^{+0.23}_{-0.26}$ & $0.0014^{+0.0008}_{-0.0006}$ & $0.77^{+0.09}_{-0.13}$ & $2.39^{+0.33}_{-0.41}$ & $1.03^{+0.63}_{-0.55}$ & $1.06^{+0.61}_{-0.53}$\\
    \rule{0pt}{3.5ex}
      CEP-0227 & $4.43^{+0.31}_{-0.35}$ & $6030^{+110}_{-110}$ & $1430^{+110}_{-100}$ & $0.81^{+0.11}_{-0.18}$ & $0.0007^{+0.0006}_{-0.0004}$ & $0.94^{+0.08}_{-0.06}$ & $1.39^{+0.89}_{-0.70}$ & $0.69^{+0.62}_{-0.43}$ & $0.82^{+0.56}_{-0.45}$\\
    \rule{0pt}{3.5ex}
      CEP-2532 & $4.06^{+0.48}_{-0.44}$ & $6340^{+140}_{-130}$ & $1150^{+100}_{-100}$ & $0.65^{+0.21}_{-0.21}$ & $0.007^{+0.006}_{-0.004}$ & $0.98^{+0.12}_{-0.10}$ & $1.55^{+0.87}_{-0.87}$ & $1.61^{+0.72}_{-0.81}$ & $1.04^{+0.69}_{-0.61}$\\
    \rule{0pt}{3.5ex}
      CEP-4506 & $3.84^{+0.39}_{-0.31}$ & $6280^{+160}_{-130}$ & $1120^{+120}_{-90}$ & $0.82^{+0.11}_{-0.17}$ & $0.0018^{+0.0014}_{-0.0009}$ & $0.94^{+0.08}_{-0.09}$ & $1.97^{+0.62}_{-0.79}$ & $1.29^{+0.85}_{-0.80}$ & $1.05^{+0.78}_{-0.58}$\\
      \hline
    \end{tabular}
  \caption{Constraints on model parameters for each Cepheid.}
  \label{tab:results}
  \end{center}
\end{table*}

The values of $M$, $T$ and $L$ inferred by Ref.~\cite{Araucaria} (along with $G_\text{LMC}=G_\text{SS}$) are shown by the dashed grey lines in Fig.~\ref{fig:posteriors}. We find our inferences to be consistent within $2\sigma$ with the results of that work (combining the two sets of uncertainties in quadrature) for all stars except CEP-1812, for which we prefer a significantly higher mass. Although we marginalise fully over all parameters in deriving our constraints---including four parameters ($G_\text{LMC}/G_\text{SS}$, $\alpha_\text{d}$, $\alpha_\text{c}$ and $\alpha$) not considered in Ref.~\cite{Araucaria}---we find our parameter uncertainties to be similar. 
%This is likely due to our more sophisticated inference framework \hd{is Pilecki using the exact same data as us in their inference?}.
The exception is the constraint on mass, which is significantly weaker in our analysis as a direct consequence of our marginalisation over $G_\text{LMC}/G_\text{SS}$. We find typical root-mean $\chi^2$ values over the four observables of 0.3 per star at the maximum-likelihood points, indicating that we are able to fit the observables within their uncertainties.

\begin{table}
  \begin{center}
  \small\addtolength{\tabcolsep}{-8pt}
    \begin{tabular}{|l|c|c|c|c|}
      \hline
      Selection & $G_\text{LMC}/G_\text{SS}$ & $\alpha_\text{d}$ & $\alpha_\text{c}$ & $\alpha$\\ 
      \hline
    \rule{0pt}{3.5ex}
      Fiducial & $0.927^{+0.051}_{-0.037}$ & $1.57^{+0.55}_{-0.50}$ & $0.90^{+0.41}_{-0.33}$ & $0.90^{+0.36}_{-0.26}$\\
    \rule{0pt}{3.5ex}
      All stars & $0.893^{+0.025}_{-0.027}$ & $2.09^{+0.29}_{-0.35}$ & $0.90^{+0.36}_{-0.28}$ & $0.91^{+0.31}_{-0.25}$\\
      \hline
    \end{tabular}
  \caption{Joint constraints on $G_\text{LMC}/G_\text{SS}$, $\alpha_\text{d}$, $\alpha_\text{c}$, and $\alpha$, combining multiple stars. The fiducial selection excludes CEP-1812; see Secs.~\ref{sec:results} and~\ref{sec:errors}.}
  \label{tab:results_joint}
  \end{center}
\end{table}

\section{Discussion}
\label{sec:discussion}

\subsection{Implications of the results}
\label{sec:significance}

Our $5\%$ constraint on $G$ in the LMC has implications for tests of fundamental physics. Modified gravity and dark energy theories generically predict a violation of the equivalence principle whereby the strength of gravity becomes spatially variable \cite{Clifton:2011jh, Baker:2019gxo}. This is true also for some resolutions of the Hubble tension \cite{D19,D20}. Any theory that is sufficiently well-specified to predict a value for $G$ in nearby galaxies in which variable stars in DEBs have been measured can be tested using our framework. With future data (see Sec.~\ref{sec:errors}) it will be possible to search for correlations of the strength of gravity with stellar, galactic, and environmental properties in the manner predicted by extensions to general relativity. Prominent examples are screened theories in which the strength of gravity typically {\it increases} in lower density regions; since the LMC is less massive than the Milky Way disk, our preference for $G_\text{LMC}<G_\text{SS}$, although statistically insignificant, is not in line with such a modification. Theories such as $f(R)$ or DGP gravity predict up to 1/3 increase in the strength of gravity in unscreened regions (though possibly much less in the LMC, which is within the Milky Way halo and may therefore be screened). Our results already rule out a 10\% increase in $G$ at the location of the LMC Cepheids at $>$3$\sigma$ significance.

Constraining spatial variation of Newton's constant is a key goal of the cosmology community. This is typically achieved through the parametrisation $G_\text{matter}-G_\text{light}$ (otherwise known as $\mu-\Sigma$), where the first parameter modifies the amplitude of Poisson's equation for the Newtonian potential $\phi$ governing the motion of non-relativistic matter, and the latter modifies Poisson's equation for the lensing potential $\phi+\psi$ governing the motion of light. Large scale structure probes afford constraints on $G_\text{matter}$ at the $20-50\%$ level \cite{BOSS_MG, Planck_MG, Joudaki_MG, DES_MGconstraints, Ferte_MG} for standard modified gravity parametrisations allowing for redshift and scale dependence in $G$. This approach analyses Fourier modes of the density and lensing fields at $0.3 \lesssim z \lesssim 1$ on scales $k \lesssim 1 h$/Mpc. By analysing specific low-redshift objects our method for constraining $G_\text{matter}$ has different systematics to these, besides being independent of cosmology and requiring no assumption for the form of $G$ as a function of length scale and time. Our measurement on the scale of 50 kpc---the distance to the LMC---helps to fill the desert between tests of $G$ at AU scales in the Solar System and $\gtrsim$10 Mpc scales in cosmology \cite{Baker:2014zba}. Further, the cosmological $G_{\rm matter}$ is only defined within the framework of linear perturbation theory.

Our inferences of the semi-analytic parameters describing hydrodynamics in stars are important for stellar astronomy. Past studies have constrained these parameters by individually varying them to achieve consistency between a fiducial model and observations (e.g. \cite{2009MNRAS.396.2194B}). For the first time we have constrained them jointly with the stellar parameters, allowing the degeneracies to be properly mapped out. Our constraints are not significantly broadened by marginalisation over $G_\text{LMC}/G_\text{SS}$ due to the lack of degeneracy between those parameters, and they will be of use for any study that requires the hydrodynamical parameters as inputs. We find evidence for $\alpha_\text{d}$ (turbulent dissipation) slightly larger than the fiducial value of 1 in Ref.~\cite{MESA_RSP}, $\sim$1.5$-2$, and $\alpha$ (mixing length parameter) slightly smaller, $\sim$0.9.

The mixing length parameter $\alpha$ in particular is a critical input for stellar modelling, and represents a major source of uncertainty in theoretical predictions. Traditionally, it has been measured by fitting numerical models to solar observations---with the most recent result implying $\alpha=1.83$ \cite{2015A&A...573A..89M}---and is then assumed to be the same for other objects. This assumption has however been challenged by Ref.~\cite{2017ApJ...840...17T}, who find evidence for a metallicity-dependence from astroseismic studies of giant stars. Our fiducial result of $\alpha=0.90^{+0.36}_{-0.26}$ is inconsistent with 1.83 at $2.6\sigma$, possibly lending credence to the hypothesis of metallicity dependence. Understanding whether $\alpha$ is indeed constant---and devising methods for measuring it in a wide variety of objects---is of paramount importance for upcoming missions such as the Transiting Exoplanet Survey Satellite (TESS) \cite{TESS}, for which the dominant source of uncertainty will be stellar modelling. As with $G_\text{LMC}/G_\text{SS}$, our constraints will strengthen significantly as more data is brought to bear, and variations with stellar properties may also readily be investigated. 

\subsection{Systematic uncertainties and future work}
\label{sec:errors}

Previously, the dominant source of systematic uncertainty in inferences of this type was the impact of nuisance parameters in the stellar modelling. We have removed this here by directly marginalising over the three most important ones, $\alpha_\text{d}$, $\alpha_\text{c}$ and $\alpha$. Nevertheless there remain significant potential systematics in our inference. These include the effects of possible variations in i) the orbital parameters of the Cepheids around the best-fit values of Ref.~\cite{Araucaria}, ii) additional \texttt{MESA} parameters describing the hydrodynamics of stellar structure, which we set to the fiducial values of Ref.~\cite{MESA_RSP}, iii) the atmosphere models that we assumed as part of our calculation, and {iv) potential deviations from Eq.~\ref{eq:D_LMC} due to mass and metallicity dependence}. Ideally each of these effects would be parametrised and marginalised over, which may become possible in the future with improved theoretical understanding and a speedup of the stellar structure simulations to permit inference in a higher-dimensional parameter space. We also note that \texttt{MESA} is a 1D code: 3D effects that it is unable to capture may provide additional sources of systematic uncertainty.

We have removed one star from our fiducial analysis, CEP-1812. This is a highly unusual object in several ways \cite{1812_1, 1812_2, Araucaria}: it has significantly lower luminosity than similar Cepheids, {it is inferred to be }$>$100 Myr younger than its red giant companion, and it appears to be crossing the instability strip for the first time (see Fig. 17 in Ref.~\cite{1812_1}). This has led Ref.~\cite{1812_2} to suggest that CEP-1812 formed from the merger of two main-sequence stars that subsequently evolved across the Hertzsprung gap, giving it more in common with Anomalous than Type-I Cepheids. These factors may bias our inference. In particular, our assumption that the hydrodynamics parameters besides $\alpha_\text{d}$, $\alpha_\text{c}$ and $\alpha$---and other control parameters of \texttt{MESA}---are constant at the fiducial values may not hold, requiring them to be either re-tuned or marginalised over. In addition, a merger origin could cause CEP-1812 to pulsate between the first and second overtones, which could render our scaling relations between linear and nonlinear period and radius (Eq.~\ref{eq:correction}) unreliable since these assumed pulsation in a single mode. Fig.~\ref{fig:posteriors} shows that the posteriors we obtain based on \texttt{MESA}'s solution for its mass, luminosity and temperature are significantly discrepant with the values reported by Ref.~\cite{Araucaria}. Further work is therefore required to analyse this star reliably and hence use it to strengthen the joint constraints on $G_\text{LMC}/G_\text{SS}$, $\alpha_\text{d}$, $\alpha_\text{c}$ and $\alpha$.

Aside from addressing these issues there are a number of ways to advance the present study. It would be desirable to perform the full joint inference of all stars simultaneously, which would obviate the need for a second MCMC step in combining the constraints but make convergence of the Markov Chain significantly more challenging. It would also be preferable to use the full nonlinear $P$ and $R$ output by \texttt{MESA} rather than employing scaling relations based on the linear values; this too would require significant advances in computing power or optimisation of \texttt{MESA} to be computationally tractable. It may be possible to speed up the inference greatly using an emulator. Finally, additional data could be brought to bear. This includes the Cepheid lightcurves measured by OGLE---which would require full \texttt{MESA} simulations to model---and measurements of additional Cepheids or other types of variable star. \texttt{MESA} is capable of efficiently modelling RR Lyrae stars \cite{MESA_RSP} which are found in binaries \cite{Lyrae_binaries}, and it may also be possible to model Type-II Cepheids to the level of accuracy our inference requires.

Other galaxies within the Local Group may be targeted. One feature of the LMC that permits our precision measurement is the fact that its distance is well constrained by the TRGB method, allowing fluxes to be converted into luminosities. This removes one degree of freedom per star. Applying our framework to other galaxies with precise distance estimates (either independent of $G$ or that can be re-calibrated as a function of $G$) will yield stronger bounds. Besides the Milky Way \cite{2020AcA....70..101S} and its satellites \cite{SMC}, a particularly promising target is Andromeda. This is sufficiently close for precise measurements and contains DEBs in a wide range of gravitational environments \cite{Ribas, Vilardell, Lee}.

\section{Conclusion}
\label{sec:conc}

We have proposed a novel test of the universality of the gravitational constant $G$. Variable stars in detached eclipsing binaries permit independent inference of stellar radius $R$, luminosity $L$, pulsation period $P$, and $G$ times mass $M$ from orbital modelling.
%Variable stars in detached eclipsing binaries permit independent inference of stellar radius $R$ and $G$ times mass $M$ from orbital modelling, luminosity $L$ from the brightness and an independent distance measurement, and pulsation period $P$ from the variability of the light curve.
We have shown that the combination of this information affords strong bounds on the value of $G$ in the stars relative to that measured in the Solar System. The constraining power comes mainly from the orbital $GM$ constraint combined with the different relative sensitivity to $G$ and $M$ of the pulsation period in the full equations solved by \texttt{MESA}.
%The constraining power comes mainly from the orbital $GM$ constraint and the different relative sensitivity to $G$ and $M$ of the pulsation period.}

%Variable stars in detached eclipsing binaries are sensitive to Newton's constant through two different effects: their masses inferred from the orbital trajectories scale as $G^{-1}$ via Kepler's third law, and, independently, the period of pulsation inferred from the light curve scales as $G^{-1/2}$. Simultaneously fitting $G$ to both measurements therefore permits a direct constraint. \hd{Change this. Maybe something along the lines of: The orbital solution provides a strong constraint on $GM$. This is broken by a different sensitivity to $G$ in the bolometric corrections and radius of the variable star, allowing a direct measurement of $G/G_\text{SS}$ in each star.}

We applied this test to the Large Magellanic Cloud, which contains six Type-I Cepheids in eclipsing binaries observed by the OGLE microlensing survey. Our fiducial analysis excludes CEP-1812, which is likely an unusual first-crossing Cepheid and possibly generated by the merger of two main-sequence stars. Performing a full Bayesian analysis marginalising over the relevant properties of the remaining stars and nuisance parameters in the implementation of hydrodynamics in \texttt{MESA}, we obtain $G_{\rm LMC}/G_{\rm SS}=0.93^{+0.05}_{-0.04}$. This 5\% constraint is a direct measurement of $G$ outside the Solar System and will strengthen as further data is brought to bear. Many extensions to general relativity and cosmological models for dark energy predict that $G$ is environment-dependent, and our method may be used to probe them directly. 

Our framework also provides a way to constrain stellar properties, with all relevant parameters marginalised over, and to infer semi-analytic parameters describing stellar hydrodynamics. In particular, assuming the mixing length parameter $\alpha$ is constant over the five stars yields $\alpha=0.90^{+0.36}_{-0.26}$. We make our code publicly available.

\bigskip

\section*{Acknowledgements} 
We are grateful for conversations with Eric Baxter, Cyrille Doux, Shahab Joudaki, Jason Kumar, Danny Marfatia, Marco Raveri, David Rubin, Jennifer van Saders, Istvan Szapudi, Xerxes Tata, and Jamie Tayar. We are indebted to Bill Paxton, Radek Smolec, and the wider MESA community for answering our many questions about RSP. 

HD is supported by St John's College, Oxford, and acknowledges financial support from ERC Grant No. 693024 and the Beecroft Trust. BJ is supported in part by the US Department of Energy Grant No. DE-SC0007901 and by NASA ATP Grant No. NNH17ZDA001N. 

This work used the DiRAC Complexity system, operated by the University of Leicester IT Services, which forms part of the STFC DiRAC HPC Facility (www.dirac.ac.uk). This equipment is funded by BIS National E-Infrastructure capital grant ST/K000373/1 and STFC DiRAC Operations grant ST/K0003259/1. DiRAC is part of the National E-Infrastructure.

\begin{figure*}[!htbp] 
  %\centering
  \includegraphics[width=0.475\textwidth]{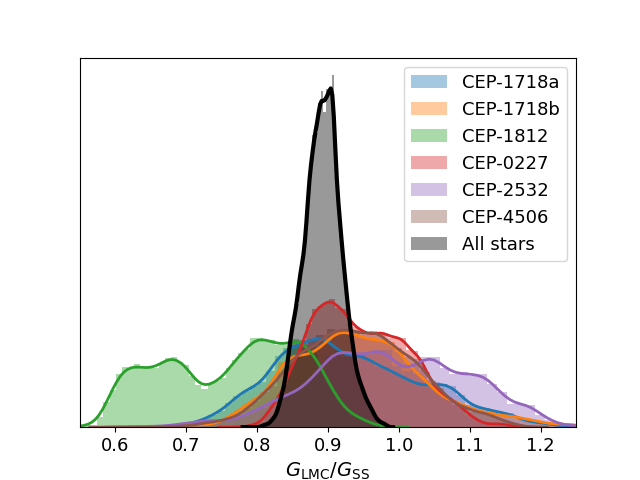}
  \includegraphics[width=0.475\textwidth]{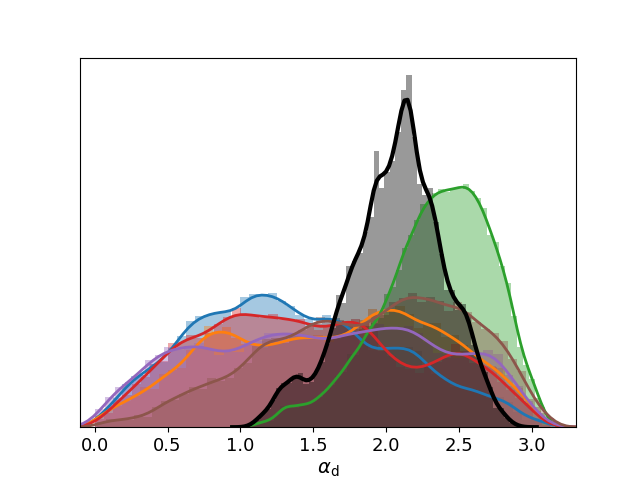}
  \includegraphics[width=0.475\textwidth]{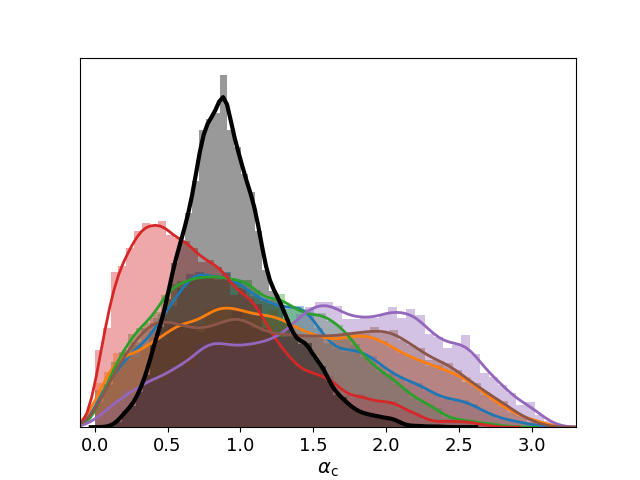}
  \includegraphics[width=0.475\textwidth]{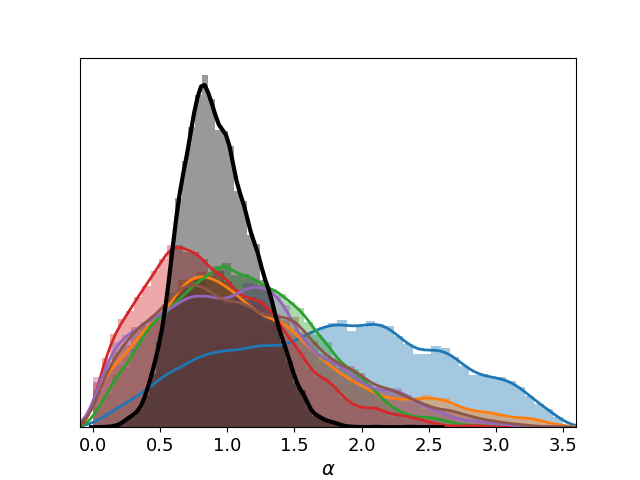}
  \caption{Combined posterior probability distributions for $G_\text{LMC}/G_\text{SS}$, $\alpha_\text{d}$, $\alpha_\text{c}$ and $\alpha$ from all 6 stars.}
  \label{fig:joint2}
\end{figure*}
\bibliography{ref_MESA}
\appendix

\section{Results including CEP-1812}
\label{sec:app}

Our fiducial analysis excludes CEP-1812 due to its tension in parameter constraints with the other stars and its peculiar nature. In Fig.~\ref{fig:joint2} we show the combined constraints on $G_\text{LMC}/G_\text{SS}$, $\alpha_\text{d}$, $\alpha_\text{c}$ and $\alpha$ when it is included. In this case we obtain a 3\% constraint on $G_\text{LMC}/G_\text{SS}$ and a $15-35\%$ constraint on the stellar hydrodynamic parameters (see second row of Table~\ref{tab:results_joint}). This result is in $4.3\sigma$ tension with $G_\text{LMC}=G_\text{SS}$, although the errorbar is biased low by the small region of overlap between the $G_\text{LMC}/G_\text{SS}$ posteriors of CEP-1812 and the other stars. Further investigation is required to determine if this star truly prefers $G_\text{LMC} < G_\text{SS}$ (see also Sec.~\ref{sec:errors}).

\end{document}